\documentclass{iopart}
\usepackage{graphicx}

\expandafter\let\csname equation*\endcsname\relax
\expandafter\let\csname endequation*\endcsname\relax

\usepackage{amsmath,amssymb,cite,mathtools}

\DeclareMathOperator\atanh{atanh}
\newcommand{\ave}[1]{\langle #1 \rangle}
\newcommand{\abs}[1]{\left| #1 \right|}
\newcommand{\pd}{\partial}
\newcommand{\diag}{\mathop{\mathrm{diag}}}

\begin{document}

\title{Inverse Ising inference with correlated samples}
\author{Benedikt Obermayer$^{1,2}$ and Erel Levine$^{1}$}

\address{$^{1}$ Department of Physics and Center for Systems Biology, Harvard University, Cambridge MA 02138, USA.}
\address{$^{2}$ Max-Delbr{\"u}ck-Center for Molecular Medicine, Berlin-Buch, Germany}
\ead{benedikt.obermayer@mdc-berlin.de, elevine@fas.harvard.edu}

\begin{abstract}
Correlations between two variables of a high-dimensional system can be indicative of an underlying interaction, but can also result from indirect effects. Inverse Ising inference is a method to distinguish one from the other. Essentially, the parameters of the least constrained statistical model are learned from the observed correlations such that direct interactions can be separated from indirect correlations. Among many other applications, this approach has been helpful for protein structure prediction, because residues which interact in the 3D structure often show correlated substitutions in a multiple sequence alignment. In this context, samples used for inference are not independent but share an evolutionary history on a phylogenetic tree. Here, we discuss the effects of correlations between samples on global inference. Such correlations could arise due to phylogeny but also via other slow dynamical processes. We present a simple analytical model to address the resulting inference biases, and develop an exact method accounting for background correlations in alignment data by combining phylogenetic modeling with an adaptive cluster expansion algorithm. We find that popular reweighting schemes are only marginally effective at removing phylogenetic bias, suggest a rescaling strategy that yields better results, and provide evidence that our conclusions carry over to the frequently used mean-field approach to the inverse Ising problem.
\end{abstract}


\maketitle

An exciting confluence of techniques from statistical physics, computer science and information theory over the last decade has yielded new methods for the study of high-dimensional interacting systems, including neuronal networks~\cite{SchneidmanNature:06}, bird flocks~\cite{BialekPNAS:12}, justices on the US supreme court~\cite{arXiv:1306.5004}, gene expression networks~\cite{LezonPNAS:06}, protein-protein interactions~\cite{WeigtPNAS:08}, transcription factor binding motifs~\cite{SantoliniarXiv:13}, HIV vaccine design~\cite{MannPCB:14}, and protein folding~\cite{MarksPO:11,MorcosPNAS:11,HopfCell:12,SulkowskaPNAS:12, MorcosPNAS:14}. Briefly, a maximum-entropy formalism~\cite{JaynesPR:57a,JaynesPR:57b} is used to infer the parameters of a Boltzmann-like probability distribution such that its first two moments coincide with the ones observed in the data. These parameters in turn can be used to distinguish direct interactions from indirect correlations. In the comparative genomics field, which is boosted by the rapid growth of sequenced genomes, such methods are used to study evolutionary correlations in protein sequences, fueled by the observation that sequence changes at one locus are frequently accompanied by compensatory changes at another locus. Assuming that this type of evolutionary constraint results from a physical interaction of the involved residues, inference of such direct correlations in multiple alignments of homologous protein sequences allows one to identify pairs of protein residues in close spatial proximity within the tertiary structure, as opposed to indirect correlations due to intermediaries~\cite{Lapedes:99}. This can be used to aid and greatly simplify computational protein structure prediction~\cite{MarksPO:11,MorcosPNAS:11,HopfCell:12,SulkowskaPNAS:12}.

Consider an alignment $\mathbf{X}$ of binary sequences from $M$ samples (e.g., species, numbered by greek indices) for $N$ sites (e.g., genomic loci, numbered by roman indices), see Fig.~\ref{fig:schema}. In a comparative genomics application, the two states $X_{\alpha i}=\pm 1$ could signify whether or not the sequence agrees with a consensus sequence, usage of a preferred or a rare codon, the presence or absence of a binding site, or any other binary observation. To obtain a description of these data with minimal prior assumptions means to infer parameters $\mathbf{h}$ and $\mathbf{J}$ of the maximum-entropy probability distribution $P(\mathbf{x}) = Z^{-1} \exp\left(\sum_i h_i x_i + \sum_{i<j} J_{ij} x_i x_j\right)$ that reproduces the observed moments $ m_i=\sum_\alpha X_{\alpha i}/M$ and $ m_{ij}=\sum_\alpha X_{\alpha i}X_{\alpha j}/M$. This is known as ``inverse Ising'' inference, and a complex global problem, since in general all inferred parameters are interdependent. 

Algorithms proposed so far include small-correlation expansions~\cite{SessakJPA:09,BialekPNAS:12}, mean-field methods~\cite{KappenANI:98,TanakaPRE:98,NguyenPRL:12}, belief propagation~\cite{MezardBook,WeigtPNAS:08}, a cluster expansion method~\cite{CoccoPRL:11,CoccoJSP:12} and logistic regression~\cite{RavikumarAOS:10,AurellPRL:12}. A common assumption is that the samples $\mathbf{X}_{\alpha}$ are independent of each other. This, however, is often not the case: for instance, aligned homologous sequences share a common evolutionary history, represented by a phylogenetic tree. Generally, the resulting correlations are always positive and give rise to biases that do not average out within the sample but lead to coherent fluctuations. Since the underlying evolutionary experiment normally cannot be repeated, there is no way to obtain a less biased estimate from independent replicates. Moreover, available sequences are usually not a fair sample of the evolutionary history, because some clades have received more attention or were more thoroughly sequenced than others (for instance, primates within mammals, or mammals within vertebrates). Alternatively, positive correlations between samples could arise when sampling too densely from a time series or Markov chain. Disregarding such correlations between samples can therefore lead to over-estimation of true correlations between sites, and significantly bias inferred parameters of the corresponding model. 

Previously it has been suggested that one could account for the redundancy in the data, e.g., due to oversampling of closely related species, by weighting the samples when calculating moments,
 $\tilde m_i=\sum_\alpha w_\alpha X_{\alpha i}$~\cite{MarksPO:11,WeigtPNAS:08,MorcosPNAS:11,CoccoPCB:13}. The weights $w_\alpha$ are chosen by heuristic methods, among them specialized weighting schemes for data from a phylogenetic tree~\cite{AltschulJMB:89,GersteinJMB:94}. However, this approach may lead to loss of information, and cannot correct for global biases. Alternatively, it was proposed that the coherent nature of phylogenetic correlations leads to a pronounced signal primarily in the first eigenvector of the observed correlations matrix~\cite{PlerouPRE:02,HalabiCell:09} and can thus be efficiently removed. Other studies (reviewed in Ref.~\cite{DutheilBB:12}) compared observed evolutionary correlations against a background expected from the phylogeny, or obtained estimates within an explicit phylogenetic model, but have not addressed the full inverse problem.

Here, we analyze inference biases due to correlated samples and propose an inverse Ising inference method to account for such correlations. Our approach is motivated by the special case of phylogenetic correlations, but our methods and conclusions also apply to between-sample correlations arising from slow dynamical processes in other contexts unrelated to biology.  The paper is organized as follows: Sec.~\ref{sec:methods} contains a definition of the problem and a detailed description of analytical and numerical methods used. The latter are not essential for a first reading of Sec.~\ref{sec:results}, which contains a discussion of our main results. Sec.~\ref{sec:discussion} discusses potential applications of our findings in the context of protein structure prediction.

\section{\label{sec:methods}Methods}

\subsection{Definition of the problem}

\begin{figure}
\centerline{\includegraphics{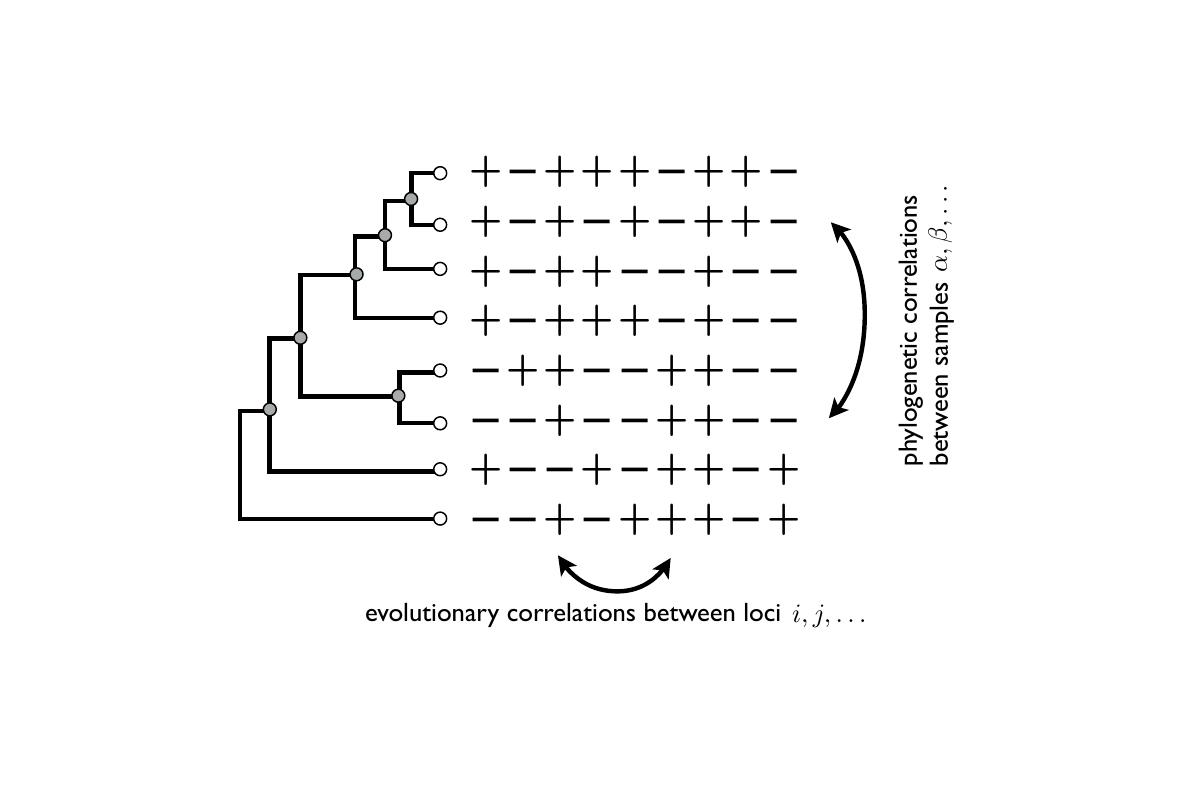}}
\caption{\label{fig:schema}Data is given in the form of an alignment $\mathbf{X}$ of $N$ loci across $M$ samples on a phylogenetic tree with $M$ external nodes (white). Data is unknown for $M-2$ ancestral nodes (grey), which are therefore integrated out. Inference of interactions between loci from observable evolutionary correlations is confounded by phylogenetic correlations between samples.}
\end{figure}

Although evolutionary dynamics does not generally occur in equilibrium, observable correlations between samples can often be well approximated by an equilibrium process. We thus assume that the entire dataset is one representative sample generated by such a known process, and estimate the remaining parameters causing deviations from expectation by maximum likelihood. Specifically, our unified framework minimizes the cross-entropy 
\begin{equation}\label{eq:cross-entropy}
\mathcal{S} = -\frac{1}{M} \ln \mathcal{P}(\mathbf{X}|\mathbf{h},\mathbf{J})
\end{equation}
\emph{of the entire alignment} with respect to the unknown parameter sets $\mathbf{h}$ and $\mathbf{J}$, where the fields $\mathbf{h}$ cause deviations of single loci from the background and the couplings $\mathbf{J}$ connect pairs of loci. This minimization is equivalent to maximizing the log-likelihood of the model given (all) the data. The $M$ species represent the leaves of an (unrooted) phylogenetic tree, with additional $M-2$ hidden (or ancestral) nodes in the interior of the tree for unknown states of common ancestors (Fig.~\ref{fig:schema}). Including these nodes into our calculation gives a larger data matrix $\mathbf{X}'$. Marginalizing over unobserved ancestral states, the probability of the data under the model reads
\begin{equation}\label{eq:probability}
\mathcal{P}(\mathbf{X}|\mathbf{h},\mathbf{J})=\frac{1}{\mathcal{Z}} \Tr' \e^{-\mathcal{H}(\mathbf{X}')}.
\end{equation}
Here, we set the energy unit $k_\text{B} T=1$, $\Tr'$ denotes a partial trace over the ancestral nodes only (i.e., the grey nodes in Fig.~\ref{fig:schema}), $\mathcal Z=\Tr \e^{-\mathcal{H}}$ is the partition function with the trace performed over all nodes, and the Hamiltonian $\mathcal{H}$ for a configuration $\mathbf{x}=(x_{\alpha i})$ is given by
\begin{equation}\label{eq:Hamiltonian}
\mathcal{H}(\mathbf{x})=\sum_i \mathcal{H}_0(\mathbf{x}^T_i)-\sum_{\alpha,i} h_i x_{\alpha i} - \sum_{\alpha,i< j} J_{ij}  x_{\alpha i}  x_{\alpha j}.
\end{equation}
Different from a standard phylogenetic approach, we model the dependencies induced by shared evolutionary history using a ``phylogenetic'' Hamiltonian 
\begin{equation}\label{eq:phylogenetic-Hamiltonian}
\mathcal{H}_0(\mathbf{x}) = -\sum_\alpha g_\alpha x_\alpha - \sum_{\alpha < \beta} K_{\alpha\beta} x_\alpha x_\beta
\end{equation}
with fields and couplings $\mathbf{g}$ and $\mathbf{K}$, respectively, where $K_{\alpha\beta}$ is nonzero only for neighbors on the phylogenetic tree, decreasing roughly with the logarithm of inverse branch length~\cite{LeuthausserJCP:86}. The fields $g_\alpha$ serve to prescribe a prior distribution on the states (e.g., beliefs about missing data or other biases for some species). By means of a reference ``background" data set $\mathbf{X}^{(0)}$, the parameters of this model ($M$ fields at the leaves of the tree and $2M-3$ couplings) can be inferred by matching the first two moments $\mu_\alpha=\ave{X^{(0)}_\alpha}$ and $\mu_{\alpha\beta}=\ave{X^{(0)}_\alpha X^{(0)}_\beta}$ between observed values and those calculated from $\mathcal{H}_0$ (see Sec.~\ref{sec:background-estimation} below). We note that this choice of phylogenetic model is based on comparable assumptions as more standard phylogenetic Markov models, and the differences lie mostly in how their parameters are interpreted (see Discussion for details). 

\subsection{A simple linear problem and local inference}

We consider first a simplified version of the problem, where the correlation structure between the $M$ species follows a linear chain rather than a tree. This model does not have hidden ancestral nodes and amounts to $N$ coupled Ising chains with fields $\mathbf{h}$, between-loci couplings $\mathbf{J}$ and between-sample coupling $K_{\alpha\beta}= K_0 \delta_{\alpha,\beta-1}$ (i.e., between neighboring rows in Fig.~\ref{fig:schema}). In this case, the partition function can be calculated using textbook transfer matrix methods. We will further restrict ourselves to the simplest case $N=2$.

Specifically, we use a system of $N=2$ Ising chains with fields $h_1$ and $h_2$, intra-chain coupling $K$ and  inter-chain coupling $J_{12}$. For large $M$, the partition function reads
\begin{equation}
\frac{1}{M}\ln \mathcal{Z} (h_1,h_2,J_{12},K) = \ln\left[ \cosh^2 K \cosh J_{12} \cosh h_1 \cosh h_2 \right] + \ln \Lambda + \mathcal{O}(M^{-1}).
\end{equation}
Here, $\Lambda$ is the largest eigenvalue of the transfer matrix, which can be written as
\begin{multline}
\diag\left(1+T,1-T,1-T,1+T\right)\times\\
\left\lbrace
\begin{bmatrix}
(1+U_1)(1+R) & (1+U_1)(1-R) \\
(1-U_1)(1-R) & (1-U_1)(1+R)
\end{bmatrix}
\otimes
\begin{bmatrix}
(1+U_2)(1+R) & (1+U_2)(1-R) \\
(1-U_2)(1-R) & (1-U_2)(1+R)
\end{bmatrix}
\right\rbrace,
\end{multline}
using $R=\tanh K$, $T=\tanh J_{12}$, and $U_{1/2}=\tanh h_{1/2}$, respectively. The eigenvalue is computed by solving
\begin{multline}\label{eq:solve_lambda}
256 R^4 (T^2-1)^2 (U_1^2-1)^2   (U_2^2-1)^2\\
-64 R^2 (R+1)^2 (T^2-1)
   (U_1^2-1) (U_2^2-1) (T U_1 U_2-1) \Lambda \\+16
   R \Big[R^2 (U_1^2 (T^2 (2
   U_2^2-1)-1)-(T^2+1) U_2^2+2)+2 R
   (T^2-1) (U_1^2 U_2^2-1)\\
   +U_1^2 (T^2
   (2 U_2^2-1)-1)-(T^2+1) U_2^2+2\Big]\Lambda^2\\-4
   (R+1)^2 (T U_1 U_2+1)\Lambda ^3 +\Lambda^4=0.
\end{multline}

\subsubsection{\label{sec:numerical-solution}Numerical solution.}

Numerical estimates $\hat h_{1/2}$ and $\hat J_{12}$ for the fields and the coupling, respectively, are calculated from the measured moments $m_1=\tfrac{1}{M}\sum_\alpha X_{\alpha,1}$, $m_2=\tfrac{1}{M}\sum_\alpha X_{\alpha,2}$ and $m_{12}=\tfrac{1}{M}\sum_\alpha X_{\alpha,1}X_{\alpha,2}$ by minimizing the entropy:
\begin{equation}
\begin{split}
\mathcal{S}(\hat h_1,\hat h_2,\hat J_{12}) =& -\hat h_1 m_1 -\hat h_2 m_2 - \hat J_{12} m_{12} +\frac{1}{M}\ln \mathcal{Z}(\hat h_1,\hat h_2,\hat J_{12},K)\\& + \mu_2 (\hat h_1^2 + \hat h_2^2) + \gamma_2 \hat J_{12}^2 +C,
\end{split}
\end{equation}
where the second line includes two regularization terms and a constant  $C=-\tfrac{1}{M}\sum_i\mathcal{H}_0(\mathbf X_i)=\tfrac{K}{M}\sum_\alpha (X_{\alpha,1}X_{\alpha+1,1}+X_{\alpha,2}X_{\alpha+1,2})$ that is ignored, such that only the partition function $\mathcal{Z}$ retains a $K$-dependence.

\subsubsection{\label{sec:analytical-solution}Analytical solution.} 

In principle, we could use the above expression for the partition function and compute the expected values $\ave{m_i}=\frac{\pd \ln\mathcal{Z}}{M\pd h_i}$ and $\ave{m_{ij}}=\frac{\pd\ln\mathcal{Z}}{M\pd J_{ij}}$ for $i,j=1,2$. Assuming that measured sample averages $m_i$ and $m_{ij}$ are representative and can be used to approximate $\ave{m_i}$ and $\ave{m_{ij}}$, respectively, these equations would then be solved to get $\hat h_i$ and $\hat J_{ij}$ with an estimated background coupling $K=\hat K$. Due to the quartic root in Eq.~\eqref{eq:solve_lambda} this is analytically impractical, even in the simple case $N=2$. We therefore treat fields and couplings independently. While it is possible (but tedious) to expand $\mathcal{Z}$ in $h_i$ and $J_{ij}$, it is much simpler to get the leading order results by solving the associated simple systems instead.

\paragraph{Inferring a field.} To first order we ignore the coupling $J_{ij}$, and only deal with one single chain of length $M$, with intra-chain coupling $K=\atanh R$ and field $h_i=\atanh U_i$. The partition function is 
\begin{equation}\label{eq:lnZ_1D}
\frac{1}{M}\ln\mathcal{Z} = \ln \frac{1+R+\sqrt{(1-R)^2+4 R U_i^2}}{\sqrt{1-R^2}\sqrt{1-U_i^2}}.
\end{equation}
From this, we compute the average magnetization as
\begin{equation}\label{eq:avem}
\ave{m_i} = \frac{\pd\ln\mathcal{Z}}{M\pd h_i}=\frac{(1+R) U_i}{\sqrt{(1-R)^2 + 4 R U_i^2}}= \frac{1+R}{1-R} h_i + \mathcal{O}(h_i^3),
\end{equation}
meaning that the inferred fields $\hat h_i$ can be calculated from observed magnetizations $m_i$ as in Eq.~\eqref{eq:simple-solution-h} below with $U_i=\tanh \hat h_i$. For small fields (and hence small magnetization), this corresponds to the expression 
\begin{equation}\label{eq:hinf}
\hat h_i = \frac{1-\hat R}{1+\hat R} m_i + \mathcal{O}(m_i^3).
\end{equation}

\paragraph{Inferring a coupling.} Here, we consider two coupled chains with intra-chain coupling $K=\atanh R$ and inter-chain coupling $J_{ij}=\atanh T_{ij}$, but no field. The partition function for this case is given by
\begin{equation}\label{eq:lnZ_2D}
\frac{1}{M}\ln\mathcal{Z}=\ln\frac{2\left[1+R^2 + \sqrt{(1-R^2)^2+ 4 T_{ij}^2 R^2}\right]}{(1-R^2)\sqrt{1-T_{ij}^2}}.
\end{equation}
This produces an average pair magnetization
\begin{equation}\label{eq:avecc}
\ave{m_{ij}} = \frac{\pd\ln\mathcal{Z}}{M\pd J_{ij}}=\frac{(1+R^2) T_{ij}}{\sqrt{(1-R^2)^2 + 4 R^2 T_{ij}^2}} = \frac{1+R^2}{1-R^2} J_{ij} + \mathcal{O}(J_{ij}^3),
\end{equation}
meaning that the coupling $J_{ij}$ is derived from observed moments $m_{ij}$ as in Eq.~\eqref{eq:simple-solution-J} below. For small $J_{ij}$, we can approximate
\begin{equation}\label{eq:Jinf}
\hat J_{ij} =  \frac{1-\hat R^2}{1+\hat R^2}m_{ij} + \mathcal{O}(m_{ij}^3).
\end{equation}
Again, ignoring the correlations between samples ($\hat K=0$) gives higher $\hat J_{ij}$ than when using a finite value.

\paragraph{Inference errors.} Since the intra-chain correlations introduce coherent fluctuations, the sample averages $m_i$ and $m_{ij}$ can be quite different from the thermodynamic averages $\ave{m_i}=\frac{\pd\ln \mathcal{Z}}{M\pd h_i}$ and $\ave {m_{ij}}=\frac{\pd \ln \mathcal{Z}}{M\pd J_{ij}}$, respectively. We can quantify the leading-order contributions to the expected inference errors $\Delta \hat h_i^2=\ave{ (\hat h_i-h_i)^2}$ and $\Delta \hat J_{ij}^2=\ave{(\hat J_{ij}-J_{ij})^2}$, by expanding in the expected fluctuations.

The expected error $\Delta h_i^2 = \langle(\hat h_i(m_i) - h_i)^2\rangle$ when using the observed value $m_i$ for inference is estimated by expanding in the difference between the error for an average observation $\ave{m_i}$ and the average error:
\begin{equation}\label{eq:inf_error_h}
\begin{split}
\langle(\hat h_i (m_i)-h_i)^2\rangle &= (\hat h_i(\ave{m_i})-h_i)^2 + \left[\langle(\hat h_i (m_i)-h_i)^2\rangle-(\hat h_i(\ave{m_i})-h_i)^2\right] \\
&\approx (\hat h_i (\ave{m_i})-h_i)^2 + \frac{\pd (\hat h_i(\ave{m_i})-h_i)^2}{2\pd \ave{m_i}^2}\left[\ave{m_i^2}-\ave{m_i}^2\right],
\end{split}
\end{equation}
where from Eq.~\eqref{eq:lnZ_1D}  we get
\begin{equation}\label{eq:avem2}
\ave{m_i^2}-\ave{m_i}^2 = \frac{\pd\ln\mathcal{Z}}{M^2\pd h_i^2}=\frac{1+R}{M(1-R)} - \frac{(1+R)(1+R(4+R))}{M(1-R)^3} h_i^2 + \mathcal{O}(h_i^4).
\end{equation}
Note that the inferred field $\hat h_i$ of Eq.~\eqref{eq:hinf} uses the \emph{assumed} intra-chain coupling $\hat K$, while the average magnetization $\ave{m_i}$ from Eq.~\eqref{eq:avem} and the fluctuation corrections $\ave{m_i^2}-\ave{m_i}^2$ from Eq.~\eqref{eq:avem2} are calculated with the \emph{actual} intra-chain coupling $K_0$ (via $R=\tanh K_0$). Combining these results in Eq.~\eqref{eq:inf_error_h} gives Eq.~\eqref{eq:inference-error-h} below.

The expected error $\langle (\hat J_{ij}-J_{ij})^2\rangle$ in the coupling is then similarly estimated by writing
\begin{equation}\label{eq:inf_error_J}
\begin{split}
\langle(\hat J_{ij} (m_{ij})-J_{ij})^2\rangle &= (\hat J_{ij}(\ave{m_{ij}})-J_{ij})^2 + \left[\langle(\hat J_{ij} (m_{ij})-J_{ij})^2\rangle-(\hat J_{ij}(\ave{m_{ij}})-J_{ij})^2\right] \\
&\approx (\hat J_{ij} (\ave{m_{ij}})-J_{ij})^2 + \frac{\pd (\hat J_{ij}(\ave{m_{ij}})-J_{ij})^2}{2\pd \ave{J_{ij}}^2}\left[\ave{m_{ij}^2}-\ave{m_{ij}}^2\right].
\end{split}
\end{equation}
We use Eq.~\eqref{eq:lnZ_2D} to get
\begin{equation}\label{eq:avecc2}
\ave{m_{ij}^2}-\ave{m_{ij}}^2 = \frac{\pd\ln\mathcal{Z}}{M^2\pd J_{ij}^2}= \frac{1+R^2}{M(1-R^2)} - \frac{(1+R^2)(1+R^2(4+R^2))}{M(1-R^2)^3} J_{ij}^2 + \mathcal{O}(J_{ij}^4).
\end{equation}
Using Eq.~\eqref{eq:avecc} and \eqref{eq:avecc2} with $R=\tanh K_0$ in Eq.~\eqref{eq:inf_error_J}, and Eq.~\eqref{eq:Jinf} with $\hat R=\tanh \hat K$ gives Eq.~\eqref{eq:inference-error-J} below.

\subsection{\label{sec:numerical-approach}Numerical approach for global inference and correlations with a tree structure}

In general, one is interested in inferring all fields and couplings simultaneously. At the same time, the correlation structure between samples is often heterogeneous. In particular, in comparative genomics applications some samples are often more similar to each other than others, reflecting the degree of shared ancestry summarized in a phylogenetic tree. Below, we detail a numerical procedure to perform global inference in the presence of between-sample correlations with a tree structure. The basic idea is to break up the system into small clusters of $n$ sites~\cite{CoccoPRL:11} and then to condense all $n$ values from one sample for each cluster into a single $2^n$-dimensional Potts spin. The interaction graph between these variables has a tree topology, and the partition function can be calculated in linear time~\cite{MezardBook}. Note that although the linear chain discussed before can be seen as a special case of the tree topology (and indeed the transfer matrix recursions are related to the belief propagation approach used below), it is much harder to derive analytical results for a tree, even when between-sample couplings are all identical: fixed points of transfer matrix or belief propagation recursions apply to \emph{bulk} spins, while observations with phylogenetic correlations come from the \emph{leaves} of the tree, and thus represent \emph{surface states} of the system.

In the following, we show how to evaluate Eq.~\eqref{eq:probability} using belief propagation where it is implicitly assumed that $N$ is a small number. The next subsection recapitulates the cluster expansion algorithm from Ref.~\cite{CoccoPRL:11,CoccoJSP:12} that can be used to systematically break down a large system into a collection of small interacting clusters.

\subsubsection{Evaluation of Eq.~\eqref{eq:probability}}

We write $\ln \mathcal{P}(\mathbf{X}|\mathbf{h},\mathbf{J})=\ln \mathcal{Z}'-\ln\mathcal{Z}$, where the restricted partition function $\mathcal{Z}'$ is computed by performing the trace only over hidden ancestral nodes, with leaf nodes fixed to observed values. We compute these two partition functions from the Bethe free energy using the same procedure~\cite{MezardBook}. In general, the Bethe free energy reads
\begin{equation}\label{eq:bethe-free-energy}
\ln \mathcal{Z} = -\sum_{(\alpha,\beta)} \sum_{\mathbf x_{\alpha},\mathbf x_{\beta}} P_2\left(H^h_{\alpha}+H^h_{\beta}+H^J_{\alpha\beta}+\ln P_2\right) 
+\sum_{\alpha}(\abs{\pd\alpha}-1)\sum_{\mathbf x_{\alpha}}P_1\left(H^h_{\alpha}+\ln P_1\right).
\end{equation}
Here, we introduced marginal distributions $P_2(\mathbf x_\alpha,\mathbf x_\beta)$ and $P_1(\mathbf x_\alpha)$, and re-organized terms of the Hamiltonian as follows: $H^h_\alpha=-\sum_i (\hat g_\alpha+h_i) x_{\alpha i} - \sum_{i<j} J_{ij}  x_{\alpha i}  x_{\alpha j}$ comes from the effective Potts field for node $\alpha$ and $H^J_{\alpha\beta}=-\sum_i \hat K_{\alpha\beta} x_{\alpha i} x_{\beta i}$ stems from the Potts coupling between two nodes. The first term in Eq.~\eqref{eq:bethe-free-energy} sums over values of neighboring nodes $(\alpha,\beta)$ and the second term runs over single nodes $\alpha$ weighted by the number $\abs{\pd\alpha}$ of neighbors. The marginal distribution $P_1(\mathbf x_\alpha)$ of a single Potts variable $\mathbf x_\alpha$ at node $\alpha$ is given by:
\begin{equation}\label{eq:node-marginal}
P_1(\mathbf x_{\alpha}) \simeq \e^{-H^h_{\alpha}} \prod_{\beta\in\pd\alpha}\sum_{\mathbf x_{\beta}} \e^{-H^{J}_{\alpha\beta}}P_{\beta \to\alpha}(\mathbf x_{\beta}),
\end{equation}
where $\simeq$ means equality up to normalization ($\sum_{\mathbf x_\alpha}P_1(\mathbf x_\alpha)=1$) and the product includes all neighbors $\pd \alpha$ of node $\alpha$. The distribution $P_2(\mathbf x_\alpha,\mathbf x_\beta)$ for a pair of neighboring nodes reads
\begin{equation}\label{eq:pair-marginal}
P_2(\mathbf x_\alpha,\mathbf x_\beta)\simeq P_{\alpha\to\beta}(\mathbf x_\alpha) \e^{-H^J_{\alpha\beta}}P_{\beta\to\alpha}(\mathbf x_\beta).
\end{equation}
Finally, both distributions use messages or beliefs $P_{\alpha\to\beta}(\mathbf x_\alpha)$, which are computed from the recursion
\begin{equation}\label{eq:message-passing}
P_{\alpha\to\beta}(\mathbf x_\alpha) \simeq \e^{-H^h_\alpha}\!\!\!\!\!\prod_{\gamma\in\pd\alpha\backslash \beta}\sum_{\mathbf x_\gamma} \e^{-H^J_{\alpha\gamma}}P_{\gamma \to\alpha}(\mathbf x_\gamma).
\end{equation}
These equations are evaluated along the tree, in one pass from the ancestral nodes outwards to the leaves, in a second pass from the leaves inwards. For the restricted partition function $\mathcal{Z}'$, we use the same method, where messages for a leaf $\alpha$ are simply fixed at the observed value $\mathbf{X}_\alpha$ by setting $P_{\alpha\to\beta}(\mathbf x_\alpha)=\delta(\mathbf x_\alpha,\mathbf{X}_\alpha)$. The entropy $\mathcal{S}=-\tfrac{1}{M}[\ln\mathcal{Z}'-\ln\mathcal{Z}]$ is then minimized with respect to the $N(N+1)/2$ parameters $\mathbf{h}$ and $\mathbf{J}$ by numerical optimization~\cite{NocedalBook}, adding $L_2$-regularization terms as prior on the coefficients. Our approach can readily be adopted to the case where only (a small number of) nonzero entries in $J_{ij}$ are to be inferred: adding $L_1$-regularization terms $\gamma_1 \|\mathbf{J}\|_1$ and using appropriate optimization methods instead enforces sparsity of the inferred matrix $\hat{\mathbf{J}}$~\cite{DonohoPNAS:03,AndrewICML:07,AurellPRL:12}.

\subsubsection{Cluster expansion for larger systems}
We expand the entropy $\mathcal{S}$ in contributions from successively larger clusters~\cite{CoccoPRL:11,CoccoJSP:12},
\begin{equation}\label{eq:entropy-expansion}
\mathcal{S}= \mathcal{S}_0 + \sum_i \Delta\mathcal{S}_i + \sum_{i,j} \Delta\mathcal{S}_{ij} + \ldots.
\end{equation}
A cluster $C=(i_1\ldots i_n)$ of $n$ spins is only included if its contribution $\Delta\mathcal{S}_C=\mathcal{S}_C - \mathcal{S}_{0,C}-\sum_{i\in C}\Delta\mathcal{S}_i-\sum_{i,j\in C}\Delta\mathcal{S}_{ij}-\ldots$ exceeds in absolute value a predefined threshold $\Theta$ after the contributions of all subclusters have been removed. Here, the entropy $\mathcal{S}_C=-\tfrac{1}{M}\ln \mathcal{P}_C(\mathbf{X}^T_C|\mathbf{h},\mathbf{J})=-\tfrac{1}{M}[\ln\mathcal{Z}_C'-\ln\mathcal{Z}_C]$ is computed from the difference in Bethe free energy Eq.~\eqref{eq:bethe-free-energy}. Larger clusters are recursively tested by merging smaller overlapping clusters. Each cluster's entropy is separately minimized with respect to its $n(n+1)/2$ associated parameters $\mathbf{h}$ and $\mathbf{J}$, and optimal parameters from overlapping clusters are summed up as described in Refs.~\cite{CoccoPRL:11,CoccoJSP:12}. The procedure terminates when no larger cluster with significant contribution to the entropy can be found. Finally, while a mean-field approximation as in Ref.~\cite{CoccoPRL:11} could be used as well (possibly including the rescaling method proposed below), for now we choose the entropy of the background model $\mathcal{S}_0=-\tfrac{1}{M}\ln\mathcal{P}(\mathbf{X}|0,0)=-\tfrac{1}{M}\sum_i \ln \mathcal{P}_i(\mathbf{X}^T_i|0,0)$ as reference point, where $\mathcal{P}_i(\mathbf{X}^T_i|0,0)=\tfrac{1}{\mathcal{Z}_0}\Tr' \e^{-\mathcal{H}_0(\mathbf{X}^T_i)}$ is the probability of a single column under the phylogenetic model Eq.~(3) with $\mathcal{Z}_0=\Tr \e^{-\mathcal{H}_0}$. Integrating a common preprocessing step, the entropy difference of single columns $\Delta\mathcal{S}_i$ or pairs of columns $\Delta\mathcal{S}_{ij}$ can then conveniently be used to decide which loci exhibit significant deviations from the background model and should be included in the inference.

\subsubsection{\label{sec:background-estimation}Background estimation}

$3M-3$ coefficients $\hat{\mathbf{g}}$ and $\hat{\mathbf{K}}$ of the phylogenetic Hamiltonian $\mathcal{H}_0$ need to be estimated from background data which plausibly evolved undisturbed by any fields $\mathbf{h}$ or couplings $\mathbf{J}$. For instance, in a protein sequence alignment one could take less conserved columns that are usually not used to infer evolutionary correlations. Given $N_0$ uncorrelated columns of such background data $\mathbf{X}^{(0)}$, one would then match marginal distributions $\pi_\alpha = \tfrac{1}{2 N_0}\sum_i (X^{(0)}_{\alpha i}+1)$ and $\pi_{\alpha\beta}=\tfrac{1}{4 N_0}\sum_i(X^{(0)}_{\alpha i}+1)(X^{(0)}_{\beta i}+1)$ to the theoretical marginals $P_1(x_\alpha=1)$ and $P_2(x_\alpha=1,x_\beta=1)=P_1(x_\alpha=1 | x_\beta=1)P_1(x_\beta=1)$ by nonlinear least squares, using Eqs.~\eqref{eq:node-marginal} and \eqref{eq:message-passing} to compute the marginals. Appropriate pseudo-counts or regularization terms should be added when estimating background parameters directly from data to avoid overfitting and reduce noise. Alternatively, if a phylogenetic Markov model for the relevant genomic regions of the species of interest is known, one could use it to calculate the marginals and then fit parameters $\mathbf{g}$ and $\mathbf{K}$. For the phylogenies of Fig.~\ref{fig:tree-creation}, we used our explicit Ising model on a perfect binary tree from which leaves were sampled, and then fitted the coefficients of $\mathcal{H}_0$ on the induced topology by exactly calculating marginals for corresponding leaves via Eq.~\eqref{eq:node-marginal}.

\section{\label{sec:results}Results}

\subsection{Analytical results for a simplified correlation structure}

To gain insight into the effect of between-sample correlations on inference, we first consider a simplified version of the problem. Instead of a branching process giving rise to a tree structure of between-sample correlations, we assume these correlations have the structure of a linear chain, as would happen if samples were taken from a time series or a Markov chain. We do not attempt to explicitly model the process that gives rise to these correlations, but assume that a linear Ising chain with intra-chain coupling $K_0$ is a sufficiently accurate description. This coupling could be estimated from background data $X^{(0)}$ for $N_0$ uncorrelated loci via $\tanh \hat K=\tfrac{1}{M N_0}\sum_{\alpha,i} X^{(0)}_{\alpha i} X^{(0)}_{(\alpha+1)i}$.

In Sec~\ref{sec:analytical-solution}, we detailed how optimal values $\hat h_i$ and $\hat J_{ij}$ can be inferred from the observed moments $m_i$ and $m_{ij}$ when treating different sites or site pairs independently:
\begin{subequations}\label{eq:simple-solution}
\begin{align}
\tanh \hat h_i &= \frac{(1-\hat R) m_i}{\sqrt{(1+\hat R)^2-4 m_i^2\hat R }}, \label{eq:simple-solution-h}\\
\tanh \hat J_{ij}&=\frac{(1-\hat R^2) m_{ij}}{\sqrt{(1+\hat R^2)^2-4 m_{ij}^2\hat R^2}}\label{eq:simple-solution-J}.
\end{align}
\end{subequations}
As expected, these estimates depend on the assumed intra-chain coupling $\hat R=\tanh \hat K$. Ignoring the phylogenetic correlations between samples (by taking $\hat K=0$ and therefore $\hat R = 0$)  would yield higher $\hat h_i$ and $\hat J_{ij}$ than when using a finite value.

Due to the coherent fluctuations induced by the between-sample correlations, the sample averages $m_i$ and $m_{ij}$ can be only poor estimators for the ensemble averages required for accurate inference. Above, the leading order contribution to the expected inference errors $\Delta \hat h_i^2=\langle (\hat h_i-h_i)^2\rangle$ and $\Delta \hat J_{ij}^2=\langle (\hat J_{ij}-J_{ij})^2\rangle$ was calculated as 
\begin{subequations}\label{eq:inference-error}
\begin{align}
\Delta \hat h_i^2 =& h_i^2 \left(\e^{-2(\hat K-K_0)}-1\right)^2 + \frac{\e^{-2(2\hat K-K_0)}}{M}, \label{eq:inference-error-h} \\
\Delta \hat J_{ij}^2 =& J_{ij}^2 \left(\frac{\cosh 2 K_0}{\cosh 2 \hat K}-1\right)^2 + \frac{\cosh 2 K_0}{M \cosh^2 2 \hat K}. \label{eq:inference-error-J}
\end{align}
\end{subequations}
The first term stems from the error made for the ``average'' configuration when neglecting or misestimating $\hat K\neq K_0$. It vanishes for perfect knowledge about the intra-chain correlations ($\hat K=K_0$), in which case the estimates for the fields $\mathbf{h}$ and couplings $\mathbf{J}$ do not incorrectly account for background correlations. The second term is a finite-size error, coming from coherent fluctuations of single finite  configurations about the average, and it therefore scales as $1/M$. Indeed, finite-size errors exist even in the uncoupled case $\mathbf h=\mathbf J=0$. While the effect of finite size fluctuations can be reduced by overestimating $\hat K$, the first term dominates for any sample of reasonable size, and the total error is minimized at (or very near) $\hat K=K_0$.

\begin{figure}
\centerline{\includegraphics{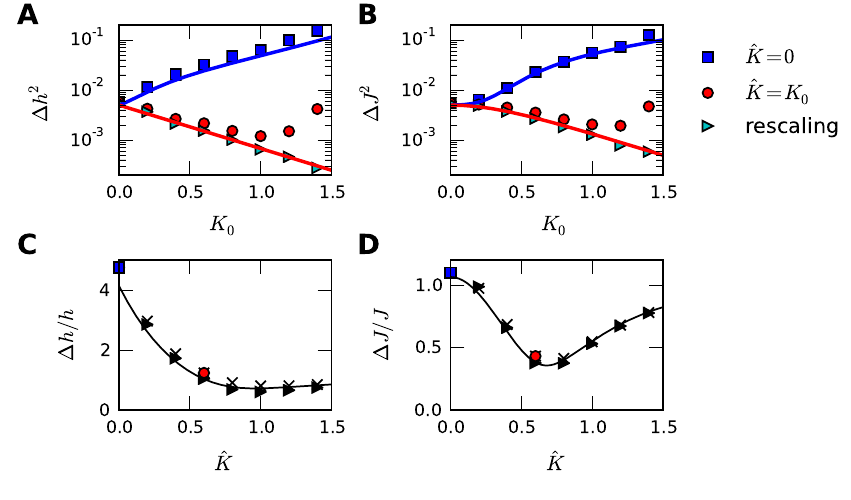}}
\caption{\label{fig:analytical-solution}Results for samples on a linear chain. Inference errors in fields (A) and coupling (B) as function of $K_0$. Errors are exponentially smaller when using the correct estimate $\hat K=K_0$ for the coupling between samples. Relative inference errors $\Delta h/h$ (C) and $\Delta J/J$ (D) as function of the assumed intra-chain coupling $\hat K$, with $K_0=0.6$ fixed (other parameters as in (A) and (B)). For finite $M$, the optimal $\hat K$ is slightly larger than $K_0$, and rescaling (triangles) gives similar results as exact inference (crosses). Solid lines are from Eqs.~\eqref{eq:inference-error}, with $\Delta h = \sqrt{\Delta h^2}$ and $\Delta J=\sqrt{\Delta J^2}$ in (C) and (D). Error bars from averaging 1000  configurations are smaller than symbol size.}
\end{figure}

To validate these results, we consider a system with $N=2$ loci and $M=200$ samples, varying the impact of correlations between the samples by increasing $K_0$. At the same time, we use an adjusted coupling $J_{12}=0.25/\cosh 2 K_0$ and fields $h_{1/2}=0.125\,\e^{-2 K_0}$ to guarantee that $m_1$, $m_2$ and $m_{12}$ are roughly independent of $K_0$, while the amplitude of coherent fluctuations increases with $K_0$. To obtain representative configurations of this system we simulated the model using a cluster Monte Carlo algorithm~\cite{WolffPRL:89,NewmanPRE:96}. Then we inferred $\hat h_{1/2}$ and $\hat J_{12}$ from each configuration (separately) via numerical minimization of the entropy $\mathcal{S}$ as in Sec.~\ref{sec:numerical-solution}. Fig.~\ref{fig:analytical-solution} shows the mean squared error in our inference across the sampled configurations as function of $K_0$ or $\hat K$, respectively, and confirms our expectations from Eq.~\eqref{eq:inference-error}. Discrepancies between theory and numerical results for higher $K_0$ are mainly due to frozen configurations which are affected by regularization. Note that \emph{relative} inference errors $\Delta \hat h/h$ and $\Delta \hat J/J$ are dominated by a global trend from our choice of adjusting fields and couplings with $K_0$, and are therefore less feasible for a comparison of results across $K_0$-values and between methods.

Intriguingly, to leading order in  $m_i$ or $m_{ij}$, the estimates in Eq.~\eqref{eq:simple-solution} become independent of $\hat K$ if rescaled values $\tilde m_i\equiv m_i\e^{-2 \hat K}$ and $\tilde m_{ij} \equiv m_{ij} / \cosh 2 \hat K$ are used. This suggests that we can simply infer $\hat{\mathbf{h}}$ and $\hat{\mathbf{J}}$ from these rescaled moments and otherwise ignore correlations between samples.  The triangles in Fig.~\ref{fig:analytical-solution} validate this procedure for our simulated data. Indeed, it works even slightly better than numerical minimization, mainly because singularities due to frozen configurations are entirely avoided, and it is also useful when $\hat K$ is not precisely estimated.

\subsection{Numerical approach for correlations with a tree structure.}

We now turn to the biologically motivated problem where the interactions between species follow a tree structure. In contrast to the above first-order analysis, we aim to infer all fields $\mathbf{h}$ and couplings $\mathbf{J}$ simultaneously. Our approach is based on maximizing the likelihood of the model parameters $\mathbf{h}$ and $\mathbf{J}$ using Eq.~\eqref{eq:probability}, with the Hamiltonian Eq.~\eqref{eq:Hamiltonian}. Parameters $\hat g_\alpha$ and $\hat K_{\alpha\beta}$ of the phylogenetic background model $\mathcal{H}_0$ (Eq.~\eqref{eq:phylogenetic-Hamiltonian}) are assumed to be known; they can be separately inferred from appropriate background data (see Sec.~\ref{sec:background-estimation}). As detailed in Sec.~\ref{sec:numerical-approach}, we can in principle evaluate Eq.~\eqref{eq:probability} by condensing all $N$ values $\mathbf{X}_\alpha$ from one species (i.e., one row in Fig.~\ref{fig:schema}) into a single $2^N$-state Potts variable. The interaction graph between these variables has a tree topology, and  the partition function can be evaluated in a time linear in $M$ using belief propagation~\cite{MezardBook}. The computational cost of this procedure grows as $M 2^{2n}$, which is obviously infeasible for systems with more than a handful of loci (i.e., larger $N$). As a solution, we combine this approach with an adaptive cluster expansion method~\cite{CoccoPRL:11,CoccoJSP:12}, in order to decompose the system into a collection of clusters of manageably small size, comprising only strongly interacting members. Fields and couplings are then inferred for each cluster separately. Briefly, the procedure starts from pairs of loci and tests for correlations by comparing the entropy (or log-likelihood) of models with and without an interaction term. This interaction is included, and the procedure is iterated to possibly expand the cluster, only if it brings a significant improvement in likelihood beyond a predefined threshold. 

\begin{figure}
\centerline{\includegraphics{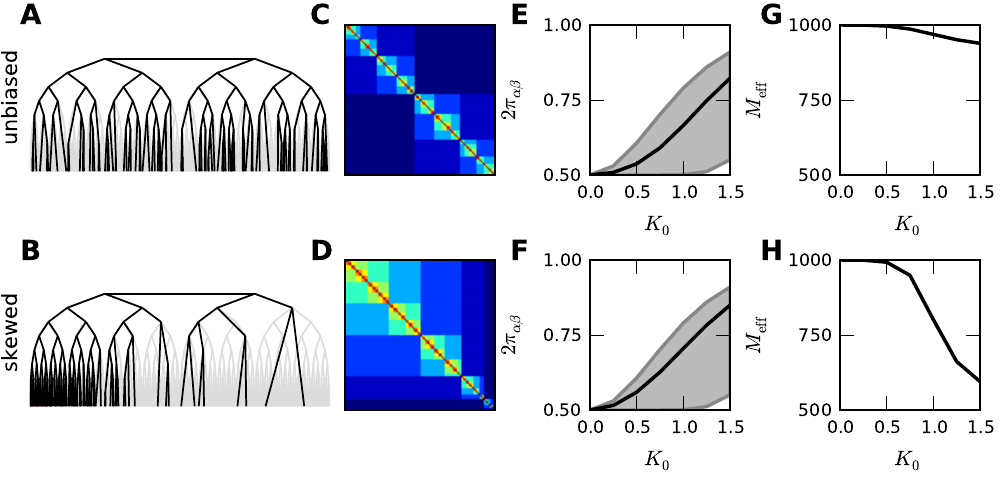}}
\caption{\label{fig:tree-creation}Creating phylogenetic trees. (A,B) Phylogenies are created by sampling $M=1000$ leaves from a perfect binary tree of 12 levels (grey; shown here with 9 levels and $M=100$) either in an unbiased (top row) or a skewed way (bottom row) to mimic sampling bias. Parameters for the new topologies (black) are inferred from phylogenetic correlations $\chi_{\alpha\beta}$ (shown as heatmaps C and D). (E,F) Range of sequence similarity $2\pi_{\alpha\beta}$ (shaded) and average similarity between most similar sequences (line). (G,H) Effective number of independent samples calculated from the information context of the weights distribution $w_\alpha$.}
\end{figure}

\paragraph{Tree generation.}
For the case of phylogenetic correlations we aim to emulate a biological problem. We create a plausible tree topology by sampling $M$ leaves from an initial perfect binary tree with homogeneous neighbor couplings $K_{\alpha\beta}\equiv K_0$ (Fig.~\ref{fig:tree-creation}(A,B)). The phylogenetic correlations between the chosen leaves are used to numerically infer the non-homogeneous parameters of the Hamiltonian $\mathcal{H}_0$ on the induced phylogeny just as would be done with real data. In terms of observables relevant in a biological context, the phylogenetic correlations are indicative of the sequence identity between two samples (the fraction of identical spins). For {\it a priori} equiprobable binary states (i.e., $g_\alpha\equiv 0$), this is calculated from $2\pi_{\alpha\beta}=\frac{1}{2}(\mu_{\alpha\beta}+1)$, which ranges from 0.5 for perfectly uncorrelated samples to 1 for perfectly correlated ones. Note that for all values $K_0 \lesssim 1$ some of the samples are actually entirely uncorrelated (see Fig.~\ref{fig:tree-creation}(E,F)).  Mimicking frequently observed sampling bias leading to a more heterogeneous correlation structure, we also create ``skewed'' topologies, where we preferentially sample leaves from one side of the tree (second row in Fig.~\ref{fig:tree-creation}). 

\begin{figure}
\centerline{\includegraphics{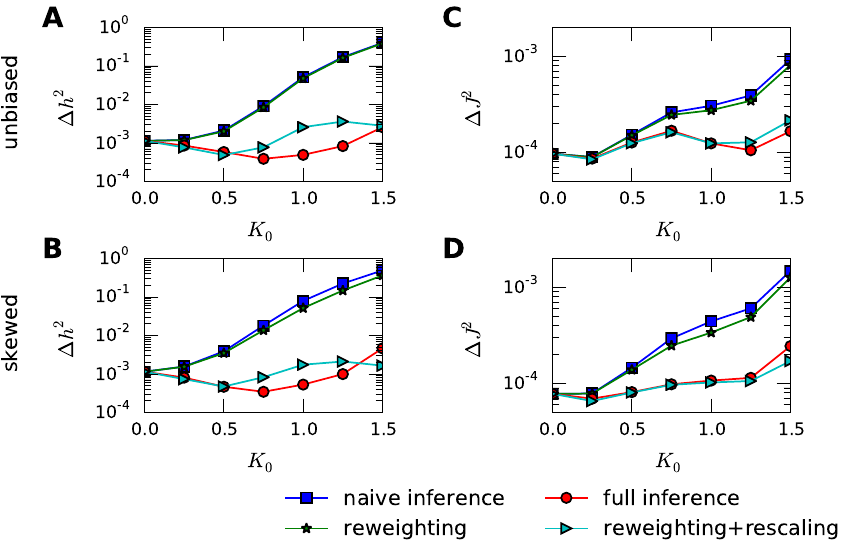}}
\caption{\label{fig:results}Results for samples on a tree. Errors in reconstructed fields $\Delta h^2$ (A,B) and couplings $\Delta J^2$ (C,D) for a system of $N=20$ loci for different inference methods as indicated. The interaction matrix $\mathbf{J}$ is sparse with $N$ entries $J_{ij}=\pm 0.25/\cosh 2 K_0$ such that no more than 3 loci are connected, and fields are uniform random numbers $\abs{h_i} \leq 0.125\, \e^{-2 K_0}$. For a tree structure this adjustment with $K_0$ does not keep the values $m_i$ and $m_{ij}$ entirely constant, but it helps to avoid frozen configurations. Error bars from averaging over $10$ configurations each for $10$ different instances of $\mathbf{h}$ and $\mathbf{J}$ are smaller than symbol size.}
\end{figure}

\begin{figure}
\centerline{\includegraphics{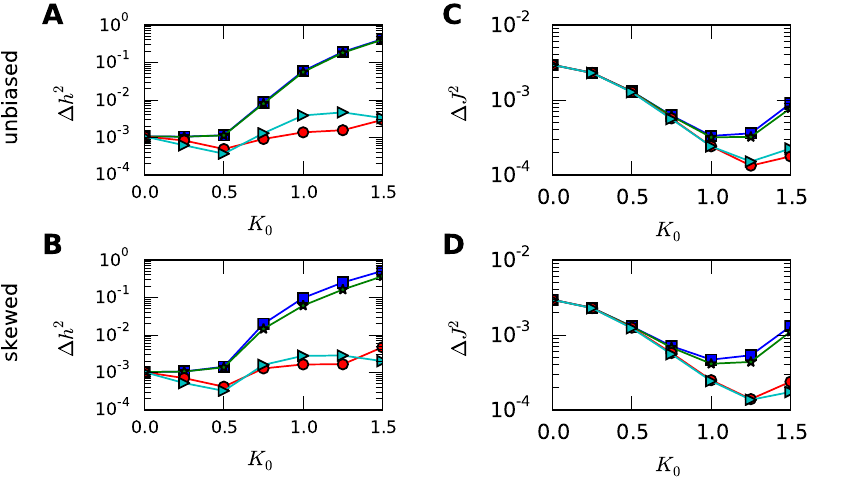}}
\caption{\label{fig:SK-results}As in Fig.~\ref{fig:results} for $N=20$, but for a Sherrington-Kirkpatrick spin glass with $h_i\equiv 0$ and $J_{ij}$ drawn from a Gaussian distribution with standard deviation $0.25/\cosh 2K_0 / \sqrt{N}$.}
\end{figure}

\paragraph{Simulation results.}
Choosing $\mathbf{h}$ and $\mathbf{J}$ as described in the legend of Figs.~\ref{fig:results} and \ref{fig:SK-results}, we generate configurations $\mathbf{X}'$ for $N$ loci on the induced phylogeny by Monte Carlo sampling~\cite{WolffPRL:89,NewmanPRE:96}. These simulated configurations are used next to reconstruct values $\hat{\mathbf{h}}$ and $\hat{\mathbf{J}}$, respectively. For a relatively simple inference problem with sparse matrix $\mathbf{J}$, Fig.~\ref{fig:results} shows the resulting mean squared errors $\Delta h^2=\frac{1}{N}\sum_i \ave{(\hat h_i-h_i)^2}$ and $\Delta J^2=\frac{2}{N(N-1)}\sum_{i< j} \ave{(\hat J_{ij}-J_{ij})^2}$, as a function of the phylogenetic coupling strength $K_0$ on the initial tree from which the leaves were sampled.  We compare our method (``full inference'') with a ``naive'' averaging using $\mathcal{H}_0=0$, where the entropy is given by the familiar expression
\begin{equation}\label{eq:entropy-uncorrelated}
\mathcal{S}_\text{uc}=\frac{1}{M}\ln \mathcal{Z}_\text{uc} - \sum_i h_i m_i - \sum_{i<j} J_{ij} m_{ij}.
\end{equation}
Here, $\mathcal{Z}_\text{uc}$ is the partition sum for the Hamiltonian Eq.~\eqref{eq:Hamiltonian} for uncorrelated samples ($\mathcal{H}_0=0$) and the averages $m_i$ and $m_{ij}$ are obtained as before by averaging over the columns of $\mathbf{X}$.

Fig.~\ref{fig:results} demonstrates that the reconstruction errors are systematically and significantly smaller using the full inference.
For better comparison between methods, we select clusters always based on differential cluster entropies for the full inference. We used a cluster threshold $\Theta=\e^{-K_0}/M$ chosen after inspection of pair entropies $\Delta\mathcal{S}_{ij}$. Otherwise, a method that is unaware of the phylogeny will always yield more clusters due to larger log-likelihood differences $\Delta \mathcal{S}$, because deviations that are actually due to phylogenetic correlations are ``explained'' by larger values for the fit parameters $\mathbf{h}$ and $\mathbf{J}$.
Similar results are obtained for a different inference problem (the Sherrington-Kirkpatrick spin glass, Fig.~\ref{fig:SK-results}). Since the trends of Fig.~\ref{fig:analytical-solution} carry over to the specific correlation structure associated with phylogenies, our analytical results are useful to understand the global inference problem.

\paragraph{Rescaling vs. reweighting.}
Previous work often used a simple reweighting approach to account for phylogenetic correlations, based on a differential weighting of samples when calculating moments from empirical observations, such that $\tilde m_i=\sum_\alpha w_{\alpha} X_{\alpha i}$ and $\tilde m_{ij} =\sum_\alpha w_{\alpha} X_{\alpha i}X_{\alpha j}$. 
For comparing to reweighing schemes, we focused on one method suitable in our context. Here, the weights $w_\alpha = \sum_\beta \chi^{-1}_{\alpha\beta}/\sum_{\gamma\delta} \chi^{-1}_{\gamma\delta}$ are calculated from the inverse of the phylogenetic correlation matrix $\chi_{\alpha\beta}=\frac{1}{4}(\mu_{\alpha\beta}-\mu_\alpha\mu_\beta)$. This gives the maximum likelihood estimate for the mean of a sample drawn from a multivariate Gaussian distribution~\cite{AltschulJMB:89}. The loss of information associated with reweighting can easily be quantified by calculating the information content $I(w)=-\sum_\alpha w_\alpha \ln w_\alpha$ of the weights distribution, and a resulting effective number of independent sequences $M_\text{eff}=\e^{I(w)}$. This reweighting scheme captures the heterogeneous structure of the phylogenetic correlations and accounts for the redundancy in the data  (Fig.~\ref{fig:tree-creation}(G,H)).

Results for ``naive'' inference with reweighting are presented as stars in Figs.~\ref{fig:results} and \ref{fig:SK-results}, and indicate significant but comparatively minor improvements, especially for the inferred couplings $\hat{\mathbf{J}}$ (see also Ref.~\cite{MorcosPNAS:11}). This implies that the specific structure of the phylogenetic tree is much less important than the overall sequence similarity in the sample. The correspondence between Figs.~\ref{fig:results} and \ref{fig:analytical-solution} therefore suggests to augment the reweighting method with a heuristic rescaling scheme, $\tilde m_i\to\tilde m_i \e^{-2 K_\text{eff}}$ and $\tilde m_{ij}\to\tilde m_{ij}/\cosh 2 K_\text{eff}$ for $i\neq j$. The effective coupling $K_\text{eff}$ serves to connect correlations on the phylogenetic tree to correlations of a linear chain. We use a well-known result for the spin-spin correlation function on a tree~\cite{MukamelPLA:74} to calculate an estimate $\tanh^2 K_\text{eff}=\frac{2}{M}\sum_\alpha\max_{\beta\neq\alpha} 2\pi_{\alpha\beta}-1$ from the average sequence similarity between most similar sequence pairs (cf.~Fig.~\ref{fig:tree-creation}(E,F)). As shown by the triangles in Figs.~\ref{fig:results} and \ref{fig:SK-results},  this simple method of globally removing phylogenetic bias significantly decreases the inference error down to the level of the full inference, even for correlations with an underlying tree structure.

\section{\label{sec:discussion}Discussion}

\subsection{The mean-field solution.}
\begin{figure}
\centerline{\includegraphics{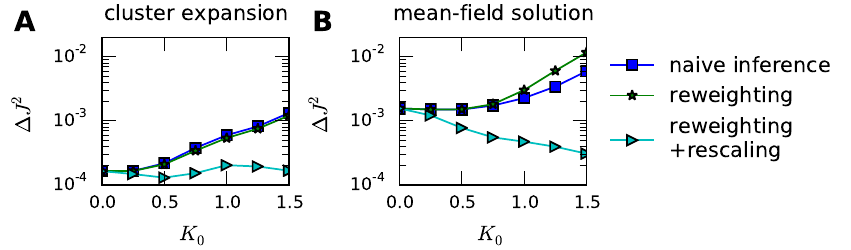}}
\caption{\label{fig:largeN}Results for a larger system with $N=50$ nodes and $M=1000$ samples from the skewed phylogeny of Fig.~\ref{fig:results}. Here, $J_{ij}=\pm 0.25/\cosh 2 K_0$ has $2N$ non-zero entries in clusters of up to 10 connected loci, and fields are as in Fig.~\ref{fig:results}. Errors in reconstructed couplings are shown for the cluster expansion (A) or the mean-field approach (B).}
\end{figure}

Recent biological applications relied on a simple mean-field approach~\cite{MarksPO:11,MorcosPNAS:11,HopfCell:12,SulkowskaPNAS:12}, where the couplings are inferred by inverting the matrix $C_{ij}=m_{ij}-m_i m_j$ of connected correlations:
\begin{equation}\label{eq:MF-solution}
\hat J_{ij} = - (C^{-1})_{ij}.
\end{equation}
To test the performance of this method in the presence of phylogenetic correlations and to compare to reweighting and rescaling schemes, we simulated data with the same phylogenies as before, but a larger system of $N=50$ loci. Fig.~\ref{fig:largeN} shows results of the inference using the cluster expansion or the mean-field method (with a pseudocount to handle insufficient variability), respectively. We did not include the full inference, since for larger systems with clusters of size $n$ the time complexity scales as $M 2^{2n}$ and additionally suffers from roundoff errors in the message passing recursions, leading to slow convergence of the minimization routines. Without the full inference as standard, we decided to include clusters based on the naive method, with the cluster threshold $\Theta=.1/M$ held fixed. Generally, the mean-field solution is less accurate than the cluster expansion, but these alternative methods follow similar trends: rescaling is very effective in removing phylogenetic biases, while reweighting is only marginally beneficial (due to a different quantitative effect of the pseudocount it actually performs worse than the naive method for larger $K_0$). 

\subsection{Connection to standard phylogenetic models}
Phylogenetic inference is usually performed using Markov models. For binary data, such models require $2M-3$ parameters (the length of each branch on the associated tree) plus one value setting the equilibrium frequency (the relative proportion of the two values). These values are fit to data using recursive algorithms largely equivalent to the ones used here~\cite{FelsensteinJME:81}. The commonly used substitution matrices also imply reversibility of the underlying stochastic process, and the assumption that the equilibrium frequency does not change along the tree. However, a simple interpretation of their parameters (e.g., branch lengths as expected substitutions per time) warrants some caution, since substitutions are not the only cause of sequence change and their rates or the relevant time unit not necessarily constant along the tree, and because other assumptions about homology and evolutionary processes enter the preparation of the alignment in the first place. More cautiously, these models can be seen as optimal descriptions of the available data within the considered space of models.

Hence we argue that our choice for describing phylogenetic correlations by means of a phylogenetic Hamiltonian is not a limiting factor, because it is merely a generalization of a Markov model to a Markov random field, allowing for different equilibrium frequencies on different leaves~\cite{WainwrightFML:07}. Alternatively, our entire approach could easily be reformulated in the language of phylogenetic models~\cite{SiepelMBE:04}, leading to similar recursions~\cite{FelsensteinJME:81}. In any case, apart from conceptual clarity and straightforward techniques for generating simulation data we believe that our non-standard formalism is advantageous under circumstances where the data is poorly fit by an explicit phylogenetic model. This could be the case due to non-uniform sequencing quality or alignability between samples, leading to an uneven distribution of gaps in the alignment. Gaps are often included as additional states, but standard Markov models prescribe constant gap frequencies along the tree~\cite{RivasPCB:08} whereas we can use different priors for each species. Further, our approach could be favorable if the data represent states of larger genomic regions, such as cis-regulatory elements, whose evolution is best described on a more coarse-grained scale. We note that exploiting the correspondence between evolutionary dynamics and Ising models has a long tradition~\cite{LeuthausserJCP:86}. A similar phylogenetic Ising model has recently been used to model HIV sequence statistics~\cite{ShekharPRE:13}.

\subsection{Inference on protein alignments}

Inverse Ising inference has found a powerful application in the prediction of residue contacts from large protein alignments (where it is often called direct coupling analysis~\cite{CoccoPCB:13,MorcosPNAS:11,SulkowskaPNAS:12,MarksPO:11,HopfCell:12}). These analyses use sequences from large protein families spanning considerable evolutionary distances, such that neutral positions in the alignment can generally be considered as independent. Still, there are typically subsets of sequences from more closely related species where this assumption is violated. In principle, our method can be readily adapted to non-binary data, corresponding to formulating the Hamiltonian Eq.~\eqref{eq:Hamiltonian} in terms of Potts variables with $\Lambda=21$ states (for 20 amino acids and a gap). In this case, we anticipate that it might be difficult to reliably estimate all associated parameters. Also, the complexity of the cluster expansion method combined with the message passing grows like $M \Lambda^{2 n}$ for a cluster of $n$ columns, which would quickly become prohibitive. Further, published methods for genomics-aided protein structure prediction~\cite{MarksPO:11,MorcosPNAS:11,HopfCell:12,SulkowskaPNAS:12} only require the identification of a small number of putative residue contacts from the top interacting pairs, and the pair ranking has been observed to be quite robust with regards to phylogenetic reweighting~\cite{WeigtPNAS:08,MorcosPNAS:11}. However, for more quantitative applications (see, e.g., Refs.~\cite{MannPCB:14,MorcosPNAS:14}), we propose the mean-field approach combined with our rescaling method as simple yet effective strategy. This mainly involves shifting measured sample averages closer to the background distribution because deviations are partially attributed to coherent fluctuations. It ameliorates problems with the proper choice of regularizers, and only requires knowledge of this background distribution and of the average sequence identity in the sample. Both can usually be reliably estimated in current sequence data sets. 

\section{Conclusions}

We presented a systematic study of inverse Ising inference for phylogenetically correlated samples, based on combining belief propagation recursions with an adaptive cluster expansion method proposed previously~\cite{CoccoPRL:11}. Here, we employed an Ising-like background model that generates the observed phylogenetic correlations. We then maximize the likelihood of interaction coefficients between different loci in adaptively chosen small clusters, given the corresponding data and the background model. Our method significantly reduces the inference error due to phylogenetic bias. Our focus here was on phylogenetic correlations between samples, but we note that such correlation may arise from slow dynamical processes in other contexts unrelated to biology. Finally, we emphasize that there might also be circumstances where biases due to phylogeny or other processes can safely be neglected, for instance if only the interaction topology (i.e., non-zero entries of $\hat{\mathbf J}$ regardless of their exact value) is of interest~\cite{DonohoPNAS:03,AndrewICML:07,AurellPRL:12}, but this question warrants further research.

Popular approaches for mitigating the effect of phylogenetic bias are based on down-weighting highly similar samples, but we show here that this has only marginal benefits. In contrast, we propose a simple rescaling of observed averages by the expected contribution attributed to excess sequence similarity, and show that it can be highly effective. Importantly, this undemanding approach is very useful even when the inference is based on simple (and computationally inexpensive) mean-field inference, which is now frequently used in the field of protein folding. 

\section*{Acknowledgments}
We thank David Nelson for discussions and Efthimios Kaxiras for critical reading of the manuscript. This work was supported by a fellowship of the German Academic Exchange Service (to BO), and by the National Science Foundation through grant MCB-1121057. 


\end{document}